\documentclass[10pt]{article}
\usepackage[sort&compress,numbers,super]{natbib}
\usepackage[version=3]{mhchem}
\usepackage[paperwidth=169mm,paperheight=239mm,total={12.7cm,21.5cm}, top=1.5cm, includeheadfoot, centering, footskip=\footskip+4mm ]{geometry}
\usepackage{mathptmx}
\usepackage{graphicx} 
\usepackage{float}
\usepackage[english]{babel}
\usepackage{charter}
\usepackage[compact]{titlesec}
\usepackage{hyperref}
\usepackage{xcolor}
\usepackage[font=small,labelfont=bf]{caption}
\usepackage[normalem]{ulem}

\begin{document}
\begin{center}
\LARGE{\textbf{Crystal-to-Crystal Transitions in Binary Mixtures of Soft Colloids}}
\end{center}

\makeatletter
\def\nlfootnote{\xdef\@thefnmark{}\@footnotetext}
\makeatother

\begin{center}
\large{Jasper N. Immink$^{\dagger,\ast}$, Maxime J. Bergman$^\ddagger$, J. J. Erik Maris$^\parallel$, Joakim Stenhammar$^\dagger$, Peter Schurtenberger$^{\dagger,\perp}$}\\
\normalsize{\textit{$^\dagger$ Division of Physical Chemistry, Lund University, Lund, Sweden. \\ $^\ddagger$ Department of Physics, University of Fribourg, Fribourg, Switzerland. \\ $^\parallel$ Inorganic Chemistry and Catalysis Group, Utrecht University, Utrecht, the Netherlands. \\ $^\perp$ Lund Institute of advanced Neutron and X-ray Science (LINXS), Lund University, Lund, Sweden. \\ $^\ast$ Tel: +46-46-2223677; E-mail: jasper.immink@fkem1.lu.se}.}
\end{center}

\section*{Abstract}
\noindent\textbf{In this article, we demonstrate a method for inducing reversible crystal-to-crystal transitions in binary mixtures of soft colloidal particles. Through a controlled decrease of salinity and increasingly dominating electrostatic interactions, a single sample is shown to reversibly organize into entropic crystals, electrostatic attraction-dominated crystals or aggregated gels, which we quantify using microscopy and image analysis. We furthermore analyze crystalline structures with bond order analysis to discern between two crystal phases. We observe the different phases using a sample holder geometry that allows both \textit{in situ} salinity control and imaging through Confocal Laser Scanning Microscopy, and apply a synthesis method producing particles with high resolvability in microscopy with control over particle size. The particle softness provides for an enhanced crystallization speed, while altering the re-entrant melting behavior as compared to hard sphere systems. This work thus provides several tools for use in the reproducible manufacture and analysis of binary colloidal crystals.}\\

\small{\noindent Keywords: binary colloidal crystals; phase transitions; colloidal particles; crystal transitions; soft colloids; tunable materials.}\\

\noindent Colloids are often used as model systems in order to study various phenomena in condensed matter physics.\cite{Frenkel2002} Particular attention has been devoted to the effects that interaction potentials have on the phase behaviour of colloids,\cite{Warren1992,Richtering1999} and to processes such as crystallization\cite{Ackerson1986, Weitz2001, Frenkel2004, Egelhaaf2011, Vanmaekelbergh2018} or the mechanisms and kinetic pathways of fluid-crystal and crystal-crystal transitions.\cite{Lekkerkerker2002, vanBlaaderen2003, Bartlett2005} In these contexts, colloids have been employed as an analogue of atomic systems, so that the relevant length and time scales are in a range that makes experimental investigation much simpler. The earliest investigations primarily used standard ``hard'' colloids such as sterically stabilized PMMA spheres.\cite{vanMegen1986,Russel1989} Later, more complex particles have been studied, to explore for instance the influence of long-range electrostatic repulsions,\cite{Lin1994,Lyonnard2000} or the effect of a short-range attraction resulting, for example, from depletion interactions in colloid-polymer mixtures.\cite{Warren1992,Zukoski2003,deKruif2003}\\ 
\noindent The majority of these investigations were performed with approximately monodisperse particles. Inspired by analogies with ``real'' atomic materials such as alloys and salts, numerous attempts have also been made to investigate binary mixtures.\cite{Pusey1992, Frenkel1993, Bartlett1994, Imhof1995, Bartlett2000, Radcliffe2005, vanBlaaderen2005, Dijkstra2006, Murray2006, Wang2018} However, investigating equilibrium structures in dense binary mixtures of hard spheres remains difficult, as nucleation and crystal growth may be very slow or hindered due to vitrification of the sample.\cite{Lowen2004,Cates2011} Moreover, the lack of simple annealing mechanisms and the resulting formation of crystals with many small domains makes a determination of the 3D crystal structure more difficult with diffraction experiments.\cite{Zhang2014,Zhang2015}\\
\noindent Here, the recent development of using soft particles as model systems\cite{Lyon2002, Yodh2014, Cloitre2014} provides us with tools that can overcome several of these hurdles. Their weak repulsions temporarily allow particle overlap, allowing particles to overcome local energy minima and promoting crystal growth speed compared to hard spheres.\cite{Hu2004,Lowen2004} Crystallization involving soft particles has been shown to possess a larger tolerance for polydispersity,\cite{Lyon2009} and many soft particle types have inherently tunable interaction potentials,\cite{Vlassopoulos2004, Breedveld2009} some as a function of externally controllable parameters such as temperature.\cite{StJohn2009} Furthermore, many soft systems can be extensively modified to tailor particle behavior or simplify analysis.\cite{Gang2003, Pelton2004, Zha2012, vonKlitzing2015} For example, modifying particles with fluorescent labels allows for direct imaging using Confocal Scanning Laser Microscopy, which provides sufficient information to determine phase behavior and specific organizations, and to quantify kinetics.\cite{Schurtenberger2015,Petukhov2017}\\
\noindent In this article, we describe several crystalline phases and a crystal-to-crystal transition. We employ oppositely charged microgel particles synthesized \textit{via} a synthesis method that emphasizes discernibility between particles in Confocal Scanning Laser Microscopy (CLSM). The softness of the particles allows for crystallization at very high concentrations\cite{Richtering1999} and the thermoresponsiveness of the particles allows for additional annealing mechanisms.\cite{Serpe2004} Furthermore, these systems allow for precise fine-tuning of volume fraction and size ratio. In the absence of electrostatic interactions, \textit{i.e.} at sufficiently high salt concentrations, the particles interact \textit{via} soft repulsive forces.\cite{Schurtenberger2013, Karg2015} We employ a sample geometry that allows \textit{in situ} salinity control during CLSM imaging, and thus a tuning of the electrostatic contributions to the interaction potential. By applying this geometry we can easily control and analyze the phase behavior of these systems as a function of electrostatic attraction strength. Our work experimentally demonstrates the existence of multiple binary soft crystal phases and how the transition between them can be controlled using \textit{in situ} control over the electrostatic interactions.

\section*{Results/discussion}
We synthesized core-shell particles with a small fluorescent core and several undyed cross-linked shells. By growing multiple shells, we achieve a thick shell with controllable size that allows good discernibility between individual particles in CLSM, while retaining the corona density distribution similar to a traditional microgel.\cite{Karg2015} Particle characteristics can be found in Table~\ref{Tab1}, with negatively charged pNIPAm particles abbreviated by pN and positively charged pNIPMAm particles by pM.\\

\begin{figure}
	\centering
	\includegraphics[width=0.65\linewidth]{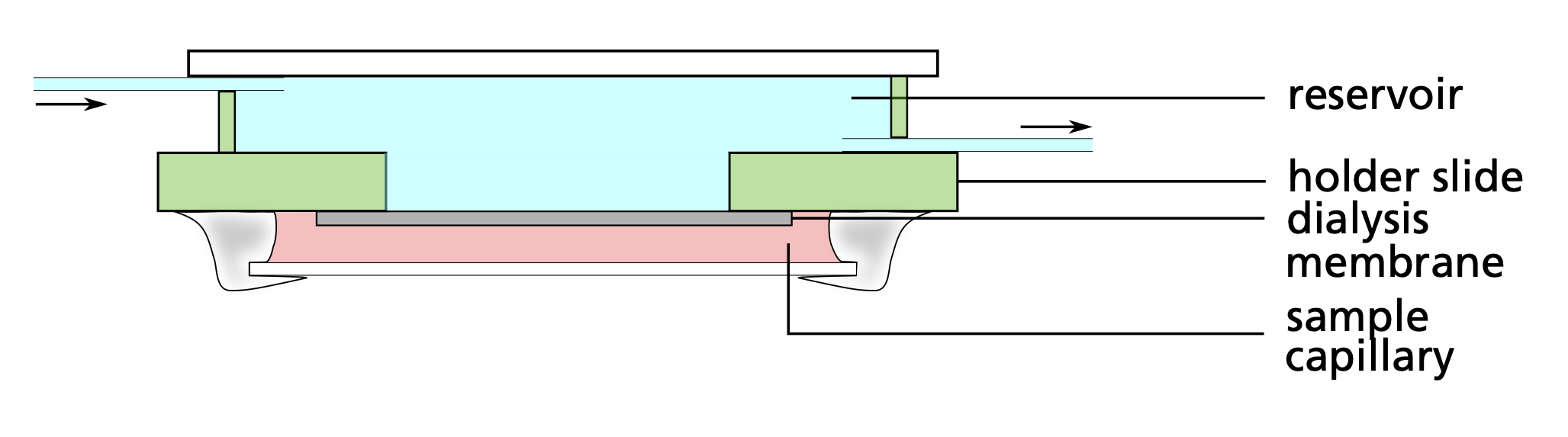}
	\caption{Cross-section of sample holder design, not to scale.}
	\label{Fig1}
\end{figure}

\noindent Samples were prepared with partial number densities of $\rho_{pN} = \rho_{pM} = 4\cdot10^{18} \text{m}^{-3}$, and were inserted in a sample cell as in Fig.~\ref{Fig1}, designed to allow controlled \textit{in situ} variation of the ionic strength during microscopical imaging. Details on design and manufacture can be found in the Materials/Methods section. The reservoir was filled with a KCl solution and left to equilibrate for at least 24 hours, kept at a constant temperature of 20~$^\circ$C.\\
\begin{figure}
	\centering
	\includegraphics[width=0.9\linewidth]{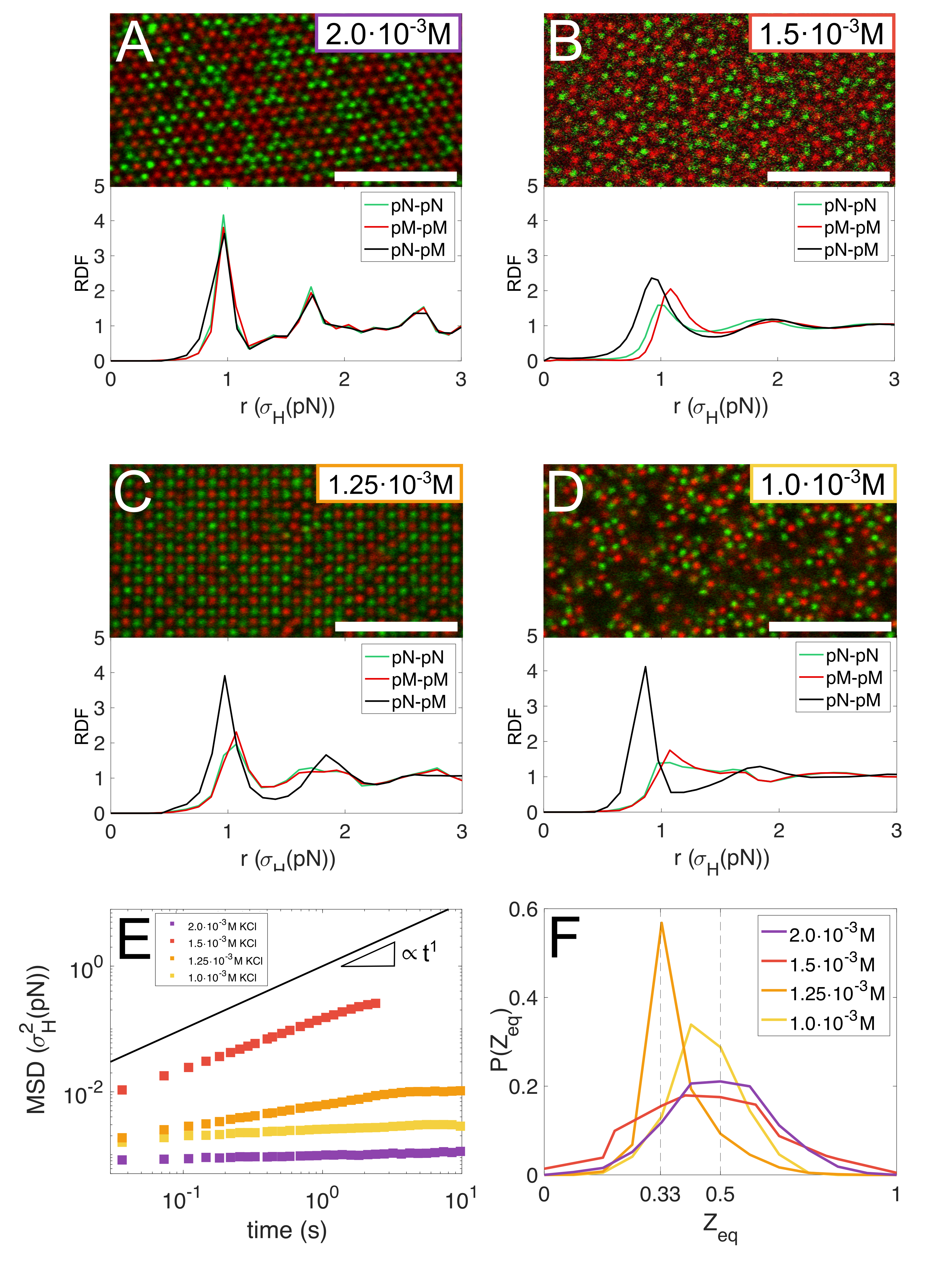}
	\caption{Analysis of one sample at different salinities. A-D) CLSM images with its corresponding radial distribution functions (RDFs) of pN-to-pN, pM-to-pM and pN-to-pM. A) $2\cdot10^{-3}$M KCl, showing entropy-driven crystallinity. B) $1.5\cdot10^{-3}$M KCl, in a fluid phase. C) $1.25\cdot10^{-3}$M KCl, showing electrostatic forces-driven crystallinity. D) Sample at $1\cdot10^{-3}$M KCl, in an electrostatically aggregated gel. E) Mean-square displacements (MSDs) of pN particles for all salinities, together with a line with a slope proportional to $t^1$ to guide the eye. The MSDs for pM are highly similar to the MSDs of pN. F) Corresponding $P(Z_{eq})$ for all salinities. Dashed lines indicate the theoretical values corresponding for AuCu (0.33) and FCC (0.5) crystal values. Each scale bar length is 10 $\mu$m.}
	\label{Fig2}
\end{figure}

\noindent In Fig.~\ref{Fig2}A-D, we show CLSM images and their 3D radial distribution functions (RDFs) for four different salinities. The calculated RDFs are split into three components, pN-to-pN, pM-to-pM, and pN-to-pM (equal to pM-to-pN). At relatively high salt concentrations ($2.0\cdot10^{-3}$M and up, Fig.~\ref{Fig2}A), charges on the particles are sufficiently screened and the system crystallizes into a hexagonal lattice due to entropic forces. The three corresponding RDFs overlap, confirming a random distribution of pN and pM particles in the lattice. It is important to note that this crystal forms despite the size ratio of 0.83 between pN and pM, highlighting the permissiveness of crystallization in soft colloids when it comes to polydispersity\cite{Lyon2009} as compared to in hard spheres.\cite{Sollich2003} Interestingly, the first peak of all RDFs of the high salt crystal, and thus its lattice spacing, appears at approximately $\sigma_H\text{(pN)}$, the hydrodynamic diameter of the smaller particle. This implies that for soft binary crystals, the smaller particles govern the lattice spacing, especially when considering that the total volume fraction is significantly above the volume fraction of a hexagonal packing of spheres: based on hydrodynamic radii and number densities, the total volume fraction is $\approx$ 1.0. This confirms earlier findings, where it has been shown that dopant-quantity microgels with large radii can adopt highly compressed conformations to fit into a crystal with a much smaller lattice spacing.\cite{Lyon2009}\\
A decrease in salinity leads to behavior increasingly dominated by electrostatic interactions, ultimately yielding aggregation and phase separation. Reducing salinity to $1.5\cdot10^{-3}$M KCl leads to melting of the entropic crystal (Fig.~\ref{Fig2}B), caused by weak electrostatic forces driving oppositely charged particles to interpenetrate, opening free space resulting in fluid behavior. It is interesting to note here that, while colloidal attractions have been shown to lead to re-entrant melting behavior,\cite{Bartsch2002} the particles' soft repulsive potential and resulting interpenetration will cause such behavior to be enhanced, as compared to incompressible hard sphere systems. Further removal of salt to $1.25\cdot10^{-3}$M KCl allows electrostatic forces to overtake entropic contributions, and causes the system to recrystallize with a specific pN-pM ordering (Fig.~\ref{Fig2}C). An even further removal of salt induces an additional increase of electrostatic attraction and yields aggregated systems ($1.0\cdot10^{-3}$M KCl, Fig.~\ref{Fig2}D). The pN-to-pM RDF reflects the increasing electrostatic attractions, with its initial peak appearing at decreasing separations $r$, reflecting the larger particle interpenetration which occurs upon increasing the electrostatic attraction strength.\\
\noindent Mean square displacements (MSDs) for pN particles at different salinities are shown in Fig.~\ref{Fig2}E, and follow the trends as described above. MSDs for pM are highly similar to those of pN and are therefore left out for clarity. At high salinities, particles are strongly caged, with apparent particle motion below the noise threshold for this analysis method; removal of salt to $1.5\cdot10^{-3}$M KCl causes cage breaking and fluidlike behavior. The slightly subdiffusive behavior is due to the high density and electrostatic attractions. At $1.25\cdot10^{-3}$M KCl, particles are once again caged, but in larger cages than at high salinities due to pN-pM interpenetration. Further removal of salt to $1.0\cdot10^{-3}$M KCl causes irreversible aggregation, where displacement is caused by cluster diffusion.\\
We define $Z_{eq}$ as the total number of nearest neighbors of equal type divided by the total number of nearest neighbors, with $P(Z_{eq})$ its probability distribution, plotted in Fig.~\ref{Fig2}F. At high salinity, $P(Z_{eq})$ is symmetric around 0.5, once again reflecting the random distribution of pN and pM particles. The weak attractions at $1.5\cdot10^{-3}$M KCl cause a preference for $Z_{eq} < 0.5$. The electrostatic crystal structure at $1.25\cdot10^{-3}$M KCl, as will be discussed later, has 4 neighbors of equal type and 8 neighbors of unequal type, reflected in the $P(Z_{eq})$ peak at $Z_{eq} = 0.33$. A further removal of salt and subsequent aggregation leads to a partial loss of preference for unequal particle types, caused by the amorphous nature of the aggregates.\\ 
Finally, we note that this behavior is fully reversible: bringing a sample from $2.0\cdot10^{-3}$M KCl to $1.0\cdot10^{-3}$M KCl and back to $2.0\cdot10^{-3}$M KCl causes an entropic crystal to melt, electrostatically aggregate, and reform into entropic crystals \textit{via} all described phases at intermediate salinities. Since the return to entropic crystals involves a transition \textit{via} a fluid phase, remixing occurs while raising salinity and a random distribution is obtained upon reaching $2.0\cdot10^{-3}$M KCl.\\

\begin{figure}
	\centering
	\includegraphics[width=0.8\linewidth]{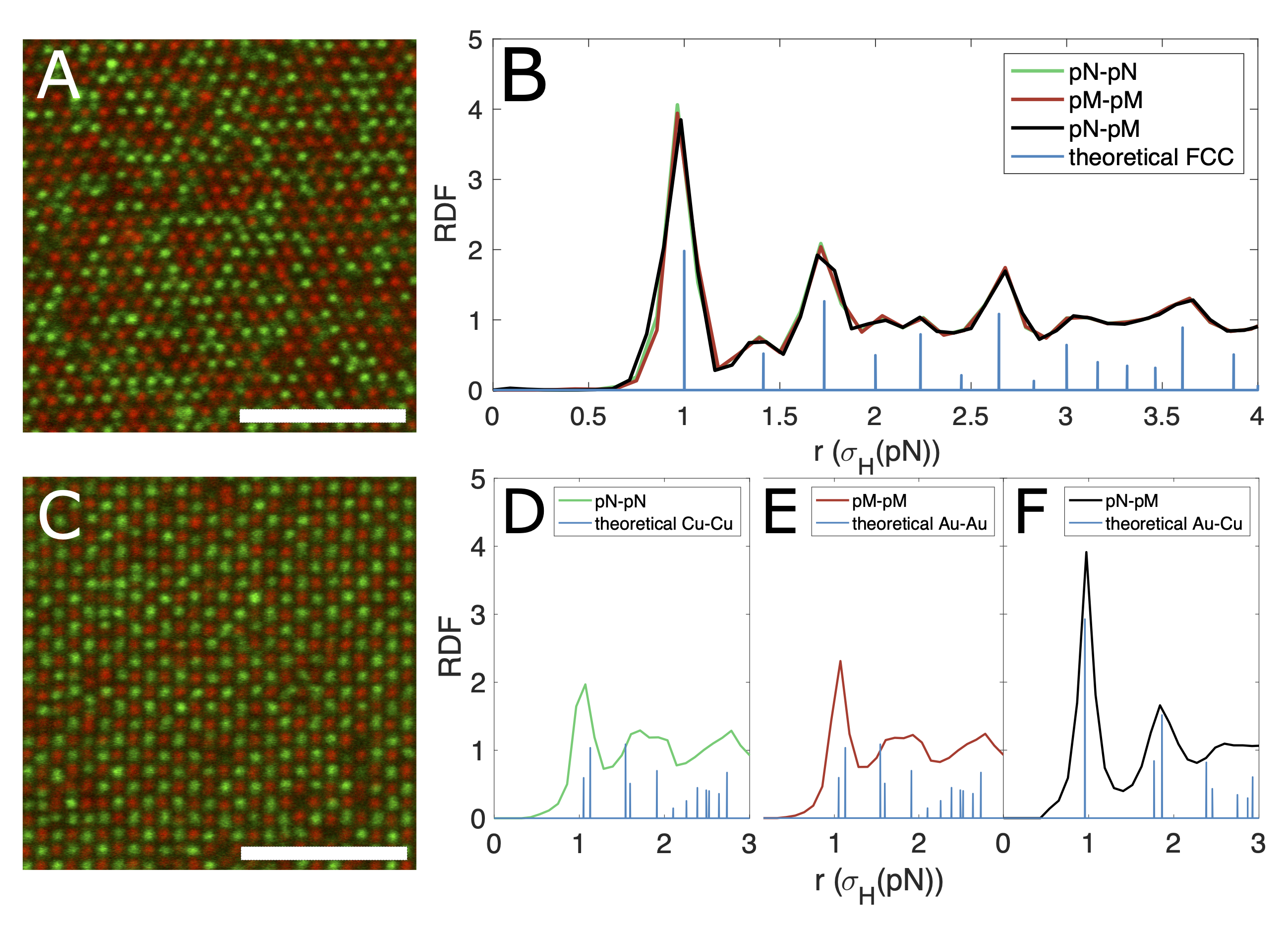}
	\caption{CLSM images of the same sample at different salinities, with its corresponding radial distribution functions (RDFs). A) CLSM image at $2\cdot10^{-3}$M KCl. B) Corresponding RDFs of pN-to-pN, pM-to-pM and pN-to-pM. The overlapping RDFs illustrates the random distribution of particles throughout the lattice. The RDF peaks are in accordance with a the positions and relative magnitudes of those in an FCC lattice (blue, rescaled magnitudes by arbitrary factor for visibility). C) CLSM image at $1.25\cdot10^{-3}$M KCl. D-F) Corresponding RDF at $1.25\cdot10^{-3}$M KCl of (D) pN-to-pN, (E) pM-to-pM and (F) pN-to-pM. Each RDF is shown with the corresponding peaks for AuCu-type crystals. The scale bar length is 10 $\mu$m.}
	\label{Fig3}
\end{figure}
\begin{figure}
	\centering
	\includegraphics[width=0.6\linewidth]{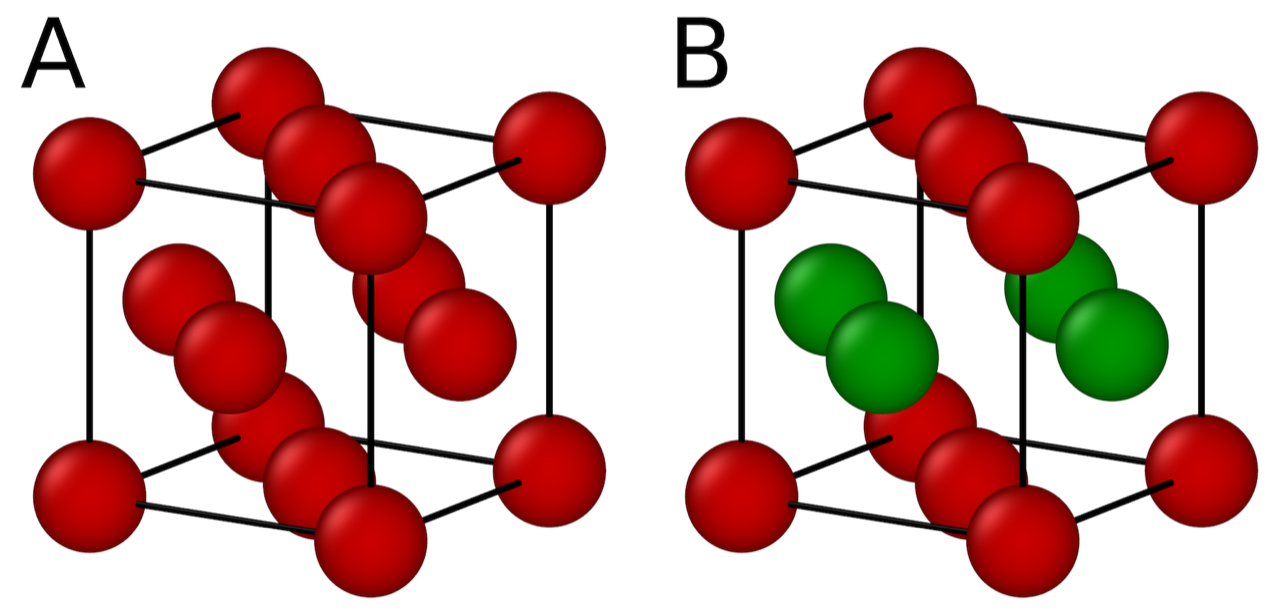}
	\caption{Unit cells of A) a single particle type FCC, and B) an AuCu crystal. The red particles represent the larger Au atoms.}
	\label{Fig4}
\end{figure}

\noindent The system is thus capable of forming two different crystal structures (Figs.~\ref{Fig2}A and \ref{Fig2}C), and a change of salinity can reversibly induce a switch between these crystal phases. For the high salt crystal at $2\cdot10^{-3}$M (shown in Figs.~\ref{Fig2}A and \ref{Fig3}A), a comparison with a theoretical FCC RDF confirms the FCC structure of the crystal, as seen in Fig.~\ref{Fig3}B. In contrast to the high salt crystal, the particles in the low salt crystal (shown in Figs.~\ref{Fig2}C and \ref{Fig3}C) are no longer randomly distributed throughout the lattice, but preferentially interact with the oppositely charged particles. Accordingly, the different RDFs seen in Figs.~\ref{Fig3}D-F, no longer overlap. These RDFs correspond well with the theoretically predicted AuCu crystal structure, with poor agreement with other AB-type binary crystal structures. A comparison between theoretical RDFs of several AB-type crystals and the experimental RDFs can be found in the Supporting Information. Note that the first peak height for the pN-to-pM RDF is significantly higher than its two equal counterparts, reflecting the strength of the electrostatic interactions that drive AuCu crystallization, overtaking entropic forces that drive FCC crystallization. In Fig.~\ref{Fig4}, unit cells of FCC-type and AuCu-type crystal structures are depicted, made using the visualization software OVITO.\cite{Ovito} The AuCu lattice parameters are identical to FCC-lattices (given identical particle sizes), but particle type distributions differ. Our observation of AuCu crystallinity corresponds with theoretical predictions for binary systems of weakly attracting, soft particles with size ratios between approximately 0.75 and 0.85.\cite{Travesset2016}\\

\noindent \noindent A common method of determining the crystal phase that a particle belongs to is by calculating local bond order parameters $q_l$.\cite{vanBlaaderen2004,Schurtenberger2015} This method analyzes the local surrounding of a particle and quantifies its symmetry, with 
$q_l$ being a measure of the local $l$-fold symmetrical order around a given particle. It is obtained by:\cite{Steinhardt1983,Dellago2008}
\begin{equation}
q_{lm}(i) = \frac{1}{Z(i)}\sum^{Z(i)}_{j=1}Y_{lm}(\textbf{r}_{ij})
\end{equation}
where $Z(i)$ is the number of nearest neighbors of particle $i$, $l$ and $m$ are integer indices, with $l \geq 0$ and $m = -l, -l + 1, ... , l - 1, l$, $Y_{lm}(\textbf{r}_{ij})$ are the spherical harmonics and $\textbf{r}_{ij}$ is a vector pointing from particle $i$ to particle $j$. $q_{lm}(i)$ is transformed into $q_l$ through:
\begin{equation}
q_l(i) = \sqrt{\frac{4 \pi}{2l + 1}\sum^l_{m = -l}|q_{lm}(i)|^2}.
\label{eq2}
\end{equation}
The FCC and AuCu lattice parameters are nearly identical, and the unit cells differ mostly by distribution of particle type; we therefore extend the analysis method by discerning three bond order parameters per particle: $q_l(\text{all})$, $q_l(\text{equal})$ and $q_l(\text{unequal})$; which are the $q_l$ calculated considering respectively all particles, only equal-type particles, and only unequal-type particles in the local surrounding. The obtained experimental values were further refined by calculating the averaged bond order parameter $\bar{q}_l$,\cite{Dellago2008} given by
\begin{equation}
\bar{q}_{lm} = \frac{1}{Z(i)}\sum^{Z(i)}_{k=0}q_{lm}(k),
\end{equation}
followed by the transformation in Eq.~\ref{eq2}. This step suppresses individual FCC-particles that randomly have AuCu-type surrounding to be designated as a particle in an AuCu phase. Order parameters for perfect FCC and AuCu crystals with identical lattice parameters are given in Table~\ref{Tab2}.
\begin{table}
\centering
\caption{Theoretical 3D local bond order parameters for FCC and AuCu crystals. $q_l$ for AuCu is split in $q_l(\text{all})$, $q_l(\text{equal})$ and $q_l(\text{unequal})$.}
\begin{tabular}{ l || l | l | l }
Crystal & $q_4$ & $q_6$ & $q_8$ \\ 
\hline
\hline
FCC & 0.19 & 0.57 & 0.40 \\
AuCu (all) & 0.19 & 0.57 & 0.40 \\
AuCu (equal) & 0.82 & 0.59 & 0.79 \\
AuCu (unequal) & 0.44 & 0.57 & 0.53 \\	
\end{tabular}
\label{Tab2}
\end{table}
\begin{figure}
	\centering
	\includegraphics[width=0.6\linewidth]{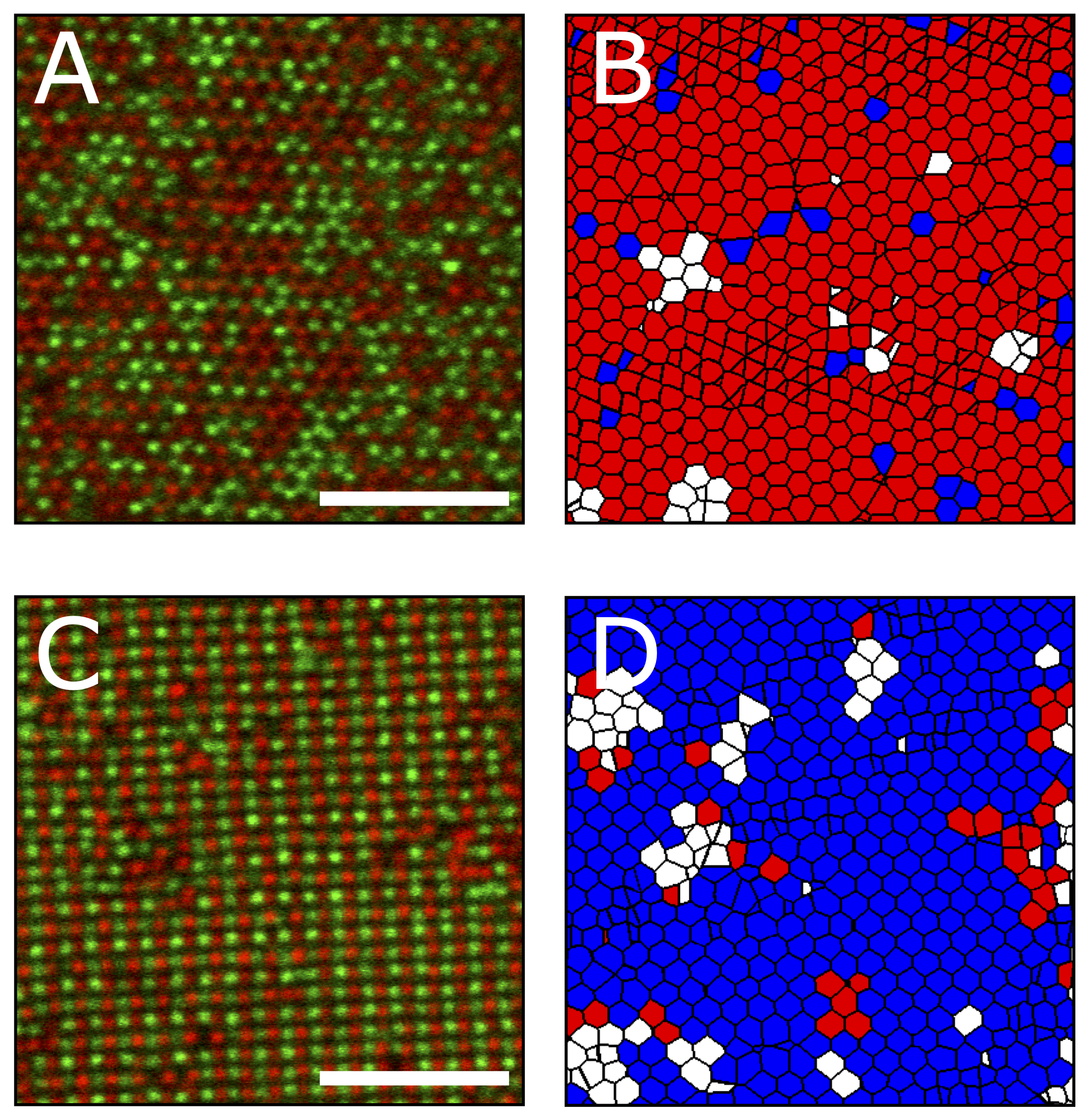}
	\caption{CLSM image slices from a 3D stack, compared with Voronoi cells colored according to their $\bar{q}_l$. A-B) FCC crystal at $2.0\cdot10^{-3}$M KCl. C-D) AuCu crystal at $1.25\cdot10^{-3}$M KCl. If $\bar{q}_4(\text{equal}) > 0.45$ or $\bar{q}_8(\text{equal}) > 0.45$, AuCu crystallinity is denoted by a blue cell; if $\bar{q}_6(\text{all}) > 0.35$, but $\bar{q}_4(\text{equal}) < 0.45$ and $\bar{q}_8(\text{equal}) < 0.45$ FCC crystallinity is denoted by a red cell; and if $\bar{q}_6(\text{all}) < 0.35$, and $\bar{q}_4(\text{equal}) < 0.45$ or $\bar{q}_8(\text{equal}) < 0.45$, the cell was left white denoting an amorphous local surrounding.}
	\label{Fig5}
\end{figure}
We calculate $\bar{q}_l$ for particles in 3D CLSM volumes and visualize this by coloring the corresponding Voronoi cells. Average $\bar{q}_l$ and standard deviations of FCC- and AuCu crystals can be found in the Supporting Information. Slices from this 3D volume are shown in Fig.~\ref{Fig5}, with Fig.~\ref{Fig5}A-B corresponding to $2.0\cdot10^{-3}$M KCl and Fig.~\ref{Fig5}C-D corresponding to $1.25\cdot10^{-3}$M KCl. While it is clear that this method works well for discerning between crystal phases, it is still sensitive to local ordering and defects. The quality of the bond order analysis benefits from the statistics provided by 3D CLSM data, and 2D microscopy imaging is not sensitive enough to discern between these two crystal structures. One point of interest is the fact that AuCu cells in the FCC lattice (Fig.~\ref{Fig5}B) are generally isolated or only a few cells, while FCC cells in the AuCu lattice (Fig.~\ref{Fig5}D) tend to be connected to other FCC cells. We attribute this to the fact that a single-particle defect in an AuCu lattice affects the local ordering for all surrounding particles, whereas several particles need to cooperate to provide the local organized surrounding of one particle in our FCC lattice. Here, one should consider that varying the ionic strength also changes the free energy landscape, as the interaction potentials change from soft repulsive to attractive, which should also influence the distribution and lifetime of defects.

\section*{Conclusions}
In this work, we have demonstrated the feasibility of binary mixtures of soft colloids for forming multiple crystal phases, including a crystal-to-crystal transition. Combining soft particles with opposite charges yields a pathway to a sequence of structural transitions that can be controlled externally \textit{via} the ionic strength of the system through a re-entrant crystal transition: \textit{i.e.} from an entropic crystal, \textit{via} an intermediate fluid regime, to an electrostatic attraction-driven crystal, and into an amorphous aggregated gel state at even stronger mutual attraction. Control over ionic strength is facilitated by the sample cell design, allowing for \textit{in situ} access to the effective interaction potential. Particles used in this study were synthesized according to a method that enhances discernibility in CLSM compared to other synthesis procedures, while allowing control over particle size during synthesis. We benefit from the softness of these particles, as they are less prone to becoming trapped in non-equilibrium arrested states at high densities, and because of their enhanced nucleation and crystal growth that facilitate \textit{in situ} studies of liquid-solid and solid-solid transitions. In addition to that, the softness and interpenetrability of the particles alters the re-entrant melting behavior as compared to hard sphere systems. Finally, we have analyzed our results using local bond order parameters adapted for discerning between systems with strong crystal similarity.\\
\noindent It is important to note that the use of these types of binary mixtures allow for an extra dimension of tunability: heating these mixtures has the effect that pN particles decrease in size, whereas pM particles are only slightly affected.\cite{Immink2019} By thus manipulating the size ratio, even more crystal phases are expected, and our system provides exceptional control over both temperature and salinity. Thus, the systems introduced here are a good candidate for observing multiple \textit{in situ} crystal-to-crystal transitions, and therefore help to bring about a deeper understanding of the kinetics and mechanisms of such transitions.

\section*{Materials/Methods}

\subsection*{Chemicals}
The monomers used were styrene (99\%, contains 4-tert-butylcatechol as polymerization inhibitor, Sigma Aldrich), N-isopropyl\-acrylamide (97\%, NIPAm, Sigma Aldrich) and N-isopropyl\-methacrylamide (97\%, NIPMAm, Sigma Aldrich). The inhibitor was removed with active basic Al$_2$O$_3$ (for chromatography, VWR chemicals). Surfactants used were sodium dodecyl sulfate (99\%, SDS, Duchefa) and cetyltrimethylammonium bromide (99\%, CTAB, Sigma Aldrich). Dyes used were Pyrromethene 546 (PM546, Exciton) and Pyrromethene 605 (PM605, Exciton). Initiators used were potassium persulfate (99\%, KPS, Sigma Aldrich) and 2,2-azobis(2-methylpropion\-amidine) dihydrochloride (97\%, V50, Sigma Aldrich). Cross-linker was N,N’-methyl\-ene-bis-acrylamide (99\%, BIS, Sigma Aldrich). 

\subsection*{Synthesis}
A core-multishell particle synthesis method was developed combining elements from previous work,\cite{Karg2015,Sprakel2015,Karg2016} yielding a small fluorescent core and an undyed soft shell, with a corona similar to traditional microgels.\cite{Karg2015} \\
\noindent pNIPAm-interlaced polystyrene cores (PS-pN) were synthesized by purging 100 mL Millipore-quality water (MQ-H$_2$O) with N$_2$ gas for 30 minutes at room temperature in a round bottom flask. 27 mL styrene was run over an Al$_2$O$_3$ column to remove polymerization inhibitor and was added to the round bottom flask together with 3 g NIPAm, 200 mg SDS and 22 mg pyrromethene 546. The mixture was stirred and purged for 45 minutes with N$_2$ at 75~$^\circ$C, while the mixture was shielded from ambient light. 50 mg KPS was dissolved in 2 mL MQ-H$_2$O, degassed, and added instantaneously to the reaction mixture. The mixture was left stirring for 24 hours, filtered through glass wool, and purified through dialysis against MQ-H$_2$O over the course of 2 weeks. Concentration was performed by centrifugation at $10^4$ g and removal of supernatant. pNIPMAm-interlaced polystyrene cores (PS-pM) cores were synthesized identically but replacing NIPAm with 3 g NIPMAm, SDS with 100 mg CTAB, pyrrhomethene 546 with 22 mg pyrrhomethene 605, and KPS with 50 mg V50. Dialysis of PS-pM particles was performed against 0.01M CTAB.\\
\noindent The pNIPAm shells were formed by purging 360 mL MQ-H$_2$O with N$_2$ gas while stirring for 30 minutes at room temperature in a round bottom flask. 2.0 g NIPAm and 138.8 mg BIS (5 mol\%) were added to the flask, and 3.56 g 15.9 wt\% suspended PS-pN was added to the mixture. The mixture was stirred and purged for 45 minutes with N$_2$ while stirring at 75~$^\circ$C, while the mixture was shielded from ambient light. 74.0 mg KPS was dissolved in 4 mL MQ-H$_2$O and added dropwise to the solution over the course of 5 minutes. After 5 more minutes, a solution of 69.4 mg BIS (2.5 mol\%) in 20 mL MQ-H$_2$O was added dropwise to the mixture over the course of 30 minutes. After 4 hours, the sample was filtered through glass wool and purified. Consecutive shell growths were performed by preparing a degassed reaction mixture with all previously grown core-shell particles suspended in a volume of 80\% of the previous reaction volume, and addition of 80\% of the previously added NIPAm, primary BIS addition (5 mol\%), secondary BIS addition (2.5 mol\%) and KPS, following identical steps. This is required to compensate for reaction yield and to prevent secondary nucleation. The final shell required no secondary BIS addition, only the addition of 5 mol\% BIS. pNIPMAm shells were grown following an identical procedure, but on pS-pM particles with 3.0 g NIPMAm instead of NIPAm, 123.3 mg BIS (5 mol\%), 61.7 mg (5 mol\%) and 74.0 mg V50 instead of KPS. Hydrodynamic diameters were determined from dynamic light scattering using a modulated 3D cross-correlation instrument (LS instruments) with a 660 nm diode-pumped laser.

\begin{table}
\centering
\caption{Particles used in experiments, with their polymer, dye type, colors, charge signs and hydrodynamic diameters $\sigma_H$ measured with dynamic light scattering at 20~$^\circ$C for the final core-shell particles and the cores only, respectively.}
{\footnotesize
\begin{tabular}{ l  l  l  l  l  l  l }
\hline
Polymer & Abbreviation & Dye & Color & Charge & $\sigma_H$(particle) [nm] & $\sigma_H$(core) [nm]\\ 
\hline
pNIPAm & pN & PM546 & Green & Negative & 280 & 73\\ 										%Real name is JICS4
pNIPMAm & pM & PM605 & Red & Positive & 336 & 103\\										%Real name is JICS7
\end{tabular}
}
\label{Tab1}
\end{table}

\subsection*{Sample Preparation}
Number densities were obtained through preparing single crystals at appropriate volume fractions, imaging a volume in CLSM, followed by particle counting. Binary microgel dispersions were prepared and mixed at 20~$^\circ$C and 10$^{-3}$M KCl.\\
\noindent Sample holders were prepared as in Fig.~\ref{Fig1}, similar to the ones described by Sato \textit{et al.}\cite{Breedveld2006} A rigid, composite material (FR-4, glass reinforced epoxy laminate) slide of dimensions 25 x 75 x 1 mm  was centrally perforated with a long, thin hole of 25 x 1 mm. The hole was covered with a dialysis membrane on one side. A capillary was created covering the dialysis membrane by placing two cover slips (5 x 50 x 0.1 mm) parallel on either side of the hole, and topped with a third cover slip (25 x 50 x 0.1 mm). The cover slips were glued in place with an air and water resistant UV glue (Thorlabs UV glue 83). The opposite side of the composite slide was fitted with a solution reservoir of a volume at least 100 times the capillary volume, with inlet and outlet tubes, and was made air and water tight using hydrophobic, malleable wax. The capillary was rinsed with MQ-H$_2$O and excess water removed before the sample was inserted. It was then sealed with UV-glue and the reservoir was filled. Measurements can be performed with a constant flow continuously refreshing the solution in the reservoir, or equilibrated at a given salinity: in this work, the latter method is used, with minimum equilibration time of 12 hours. Reversibility was tested by slowly cycling samples between low and high salinities at least three times, while checking the appearance of all states shown in Fig.~\ref{Fig2}; this was successful for at least three cycles in three separately prepared samples. All samples were placed in a 0.02\% (w/v) NaN$_3$ solution for 24 hours per week to prevent bacterial growth.

\subsection*{Imaging}
Samples were mounted on an inverted CLSM (Leica TCS SP5 tandem scanner), and imaged using a 100x/1.4 NA oil immersion objective. The microscope was mounted in an enclosure that allows for temperature control with a 0.2~$^\circ$C maximum variance using thermostated air circulation. Using standardized image analysis and particle tracking routines,\cite{Grier1996} particle center coordinates were obtained for mean square displacements, radial distribution functions (RDFs) and nearest-neighbor analyses. The uncertainty in these coordinates is approximately 14 nm, obtained by determining particle centers of immobilized polstyrene particles over time.\cite{Stradner2019} The particle geometry, with its small fluorescent core and large undyed shell, facilitates the enhanced particle center determination. Particle coordinates were determined using methods described in Ref.~\citenum{Schurtenberger2014} for at least 10 z-stacks per different salinity.

\section*{Acknowledgements}
We gratefully acknowledge financial support from the European Research Council (ERC-339678-COMPASS) and the Swedish Research Council (2015-05449 and 2018-04627).

\section*{Supporting Information}

\textbf{Supporting Information Available:} Experimental RDFs compared to theoretical AuCu RDF; Experimental RDFs compared to theoretical CsCl RDF; Experimental RDFs compared to theoretical NaCl RDF; Obtained bond order parameters for all particles at $2.0\cdot10^{-3}$M KCl; Obtained bond order parameters for all particles at $1.25\cdot10^{-3}$M KCl. This material is available free of charge \textit{via} the Internet at charge at https://pubs.acs.org/doi/10.1021/acsnano.0c03966.

\footnotesize{
\bibliographystyle{unsrt}
\bibliography{bib_cry1}
}

\end{document}